\documentclass{iaus}
\usepackage{graphicx} 

\title[Optical spectroscopy of DPVs and the case of LP Ara] 
{Optical spectroscopy of DPVs \\ and the case of LP Ara}

\author[Mennickent et al.]   
{Ronald E. Mennickent$^1$, 
Darek Graczyk$^1$, Zbigniew Ko{\l}aczkowski$^2$, Gabriela Michalska$^2$, 
Daniela Barr{\'{i}}a$^1$ \and Ewa Niemczura$^{2}$}

\affiliation{$^1$Dpto. de Astronom\'{\i}a, Universidad de Concepci\'on, Chile,   
$^2$Instytut Astronomiczny Uniwersytetu Wroc{\l}awskiego, Wroc{\l}aw, Poland
 }

\pubyear{2011}
\volume{272}  
\pagerange{1-2}
\jname{Active OB stars: structure, evolution, mass loss and critical limits}
\editors{C. Neiner, G. Wade, G. Meynet \& G. Peters, eds.}
\begin{document}

\maketitle

\begin{abstract}
We present preliminary results of our spectroscopic campaign of a group of intermediate mass interacting binaries dubbed ``Double Periodic Variables" (DPVs), characterized by orbital light curves and additional long photometric cycles recurring roughly after 33 orbital periods (Mennickent et al. 2003, 2005). They have been interpreted as interacting, semi-detached binaries showing cycles of mass loss into the interstellar medium (Mennickent et al. 2008,  Mennickent \& Ko{\l}aczkowski 2009). High resolution Balmer and helium line profiles of DPVs can be interpreted in terms of mass flows in these systems. A system solution is given for LP\,Ara, based on modeling of the ASAS V-band orbital  light curve and  the radial velocity  of  the donor star. 
\end{abstract}

\firstsection 
\section{Spectra of Galactic DPVs and report on the analysis  of LP\,Ara}

During recent years we have monitored a sample of Galactic DPVs with high resolution optical spectrographs. From the inspection of the spectral region around  H$\alpha$ and He\,I\,5875 we find in all cases evidence for blended emission or absorption profiles of complex morphology (Fig.\,1). 
The He\,I \,5875 profiles are usually broad and shallow, being the AU\,Mon He\,I 5875 profile exceptionally deep among Galactic DPVs.

LP Ara (HD\,328568, 2MASS\,J16400178-4639348, B = 10.48, B-V = 0.28) is classified as an eclipsing binary of $\beta$ Lyr type in SIMBAD (simbad.u-strasbg.fr/simbad/). Spectral types B8+[A8] and mass ratio $q$ = 0.090 were given by Svechnikov \& Kuznetsova (1990). From modeling of photometric observations made with the INTEGRAL/OMC camera, Zasche (2010) found a semidetached system with  orbital period $P_{o}$ = 8.53282038 d, $i$ = 77.1$^{o}$, $q$ = 0.2, ratio between stellar temperatures and radii $T_{1}/T_{2}$ = 1.143 and $R_{1}/R_{2}$ = 1.135 and no third light. 
The above authors
did not correct their observations for the additional long photometric cycle $P_{l}$  =  273 days reported by Michalska et al. (2009).  

We compared the spectrum taken  near the long cycle maximum at $\Phi_{o}$ = 0.96  with a grid of synthetic model spectra  in a region deployed of H\,I and He\,I lines. We find the best fit for the secondary star with the model $T_{eff}$  = 9500 K, log g = 3.0 and $v_{2}$ $\sin$i = 65 km/s. We modeled the ASAS-3 light curve and radial velocity  of LP Ara with a  Wilson-Devinney code obtaining $P_{o}$ = 8.53295 d, $T_{1}$  = 16400 K, $q$  = 0.30, $i$  = 83.9$^{o}$, orbital separation $a$  = 41.1 $R_{\odot}$,  mass funtion $f(m)$ = 5.70 $\pm$ 0.36, $M_{1}$ = 9.84 $M_{\odot}$,  $M_{2}$ = 2.98 $M_{\odot}$, $R_{1}$ = 5.3 $R_{\odot}$, $R_{2}$ = 11.6 $R_{\odot}$, $\log g_{1}$  = 4.0, $\log g_{2}$ = 2.8 and $V$-band luminosity ratio $k$ = $L_{1}/L_{2}$ = 1.50. Typical errors of derived physical parameters are $\approx$ 10-20\%. 
LP\,Ara is a double lined spectroscopic binary, however according to the present state of our analysis, only lines from the secondary component strictly follow the orbital motion. The use of ASAS-3 photometry corrected for long period changes yields a model free from systematic effects  within 5\% accuracy. If third light or additional structures do exist they contribute below 5\% to the total orbital light.

\section{Conclusions}

DPV H$\alpha$ profiles are complex and usually show asymmetric absorption/emission features varying with the orbital period as well as with the long cycle. This fact suggests that {\it often} the line emission region is not disc-like, but more as an irregular structure, a fact  already noted for V\,393 Sco (Mennickent et al. 2010). The photometric regularity of DPVs (Michalska et al. 2009) place them apart from active Algols (W Serpentids). 
The rotational velocities of emitting material in some DPVs are much larger than expected for Keplerian orbits around B-type primaries.  The system parameters for LP Ara fit the global scheme of low mass ratios found in other DPVs, e.g. OGLELMC-SC8-125836 and V\,393\,Scorpii  (Mennickent et al.  2008, 2010) and  AU\,Mon (Desmet et al. 2010).  \\

\begin{figure}[b]
\begin{center}
  \includegraphics[width=4.6in]{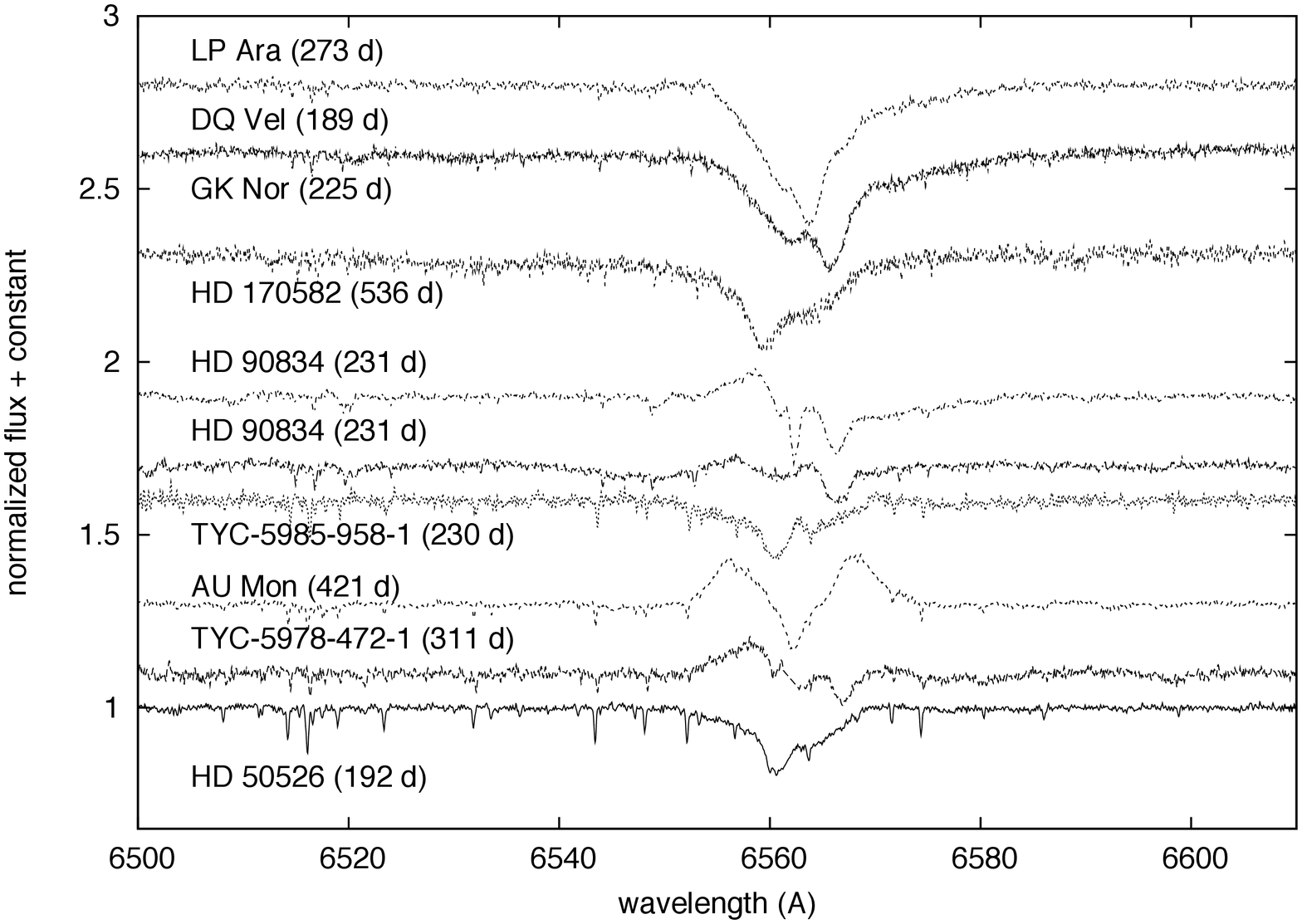} 
 \caption{Spectra of Galactic DPVs around H$\alpha$ at randomly selected orbital phases. The long period, found by us from a study of ASAS-3 ligth curves, is given in parenthesis. Two spectra of HD\,90834 illustrate line profile variability. Sharp absorption features are telluric lines. 
}
   \label{fig1}
\end{center}
\end{figure}

\end{document}